\RequirePackage{fix-cm}
\documentclass[twocolumn]{svjour3}          
\smartqed  
\usepackage{graphicx}
\usepackage{bm}
\journalname{jcamd}

\usepackage[numbers]{natbib}
\usepackage{booktabs}
\usepackage{comment,footnote}
\usepackage{multirow}
\usepackage{multicol}

\begin{document}

\title{Application of the Alchemical Transfer and Potential of Mean Force Methods to the SAMPL8 Host-Guest Blinded Challenge}

\titlerunning{SAMPL8 with ATM and PMF}        

\author{Solmaz Azimi    \and
        Joe Z. Wu \and
        Sheenam Khuttan \and
        Tom Kurtzman \and
        Nanjie Deng \and
        Emilio Gallicchio
}

\institute{
S. Azimi \at
              Department of Chemistry, Brooklyn College of the City University of New York \\
              PhD Program in Biochemistry, Graduate Center of the City University of New York
              \and
J. Z. Wu \at
              Department of Chemistry, Brooklyn College of the City University of New York \\
              PhD Program in Chemistry, Graduate Center of the City University of New York
             \and
S. Khuttan \at
              Department of Chemistry, Brooklyn College of the City University of New York \\
              PhD Program in Biochemistry, Graduate Center of the City University of New York
           \and
T. Kurtzman \at
              Department of Chemistry, Lehman College of the City University of New York \\
              PhD Program in Chemistry, Graduate Center of the City University of New York \\
              PhD Program in Biochemistry, Graduate Center of the City University of New York
             \and
N. Deng \at
             Department of Chemistry and Physical Sciences, Pace University, New York, New York
             \and
E. Gallicchio \at
              Department of Chemistry, Brooklyn College of the City University of New York \\
              PhD Program in Chemistry, Graduate Center of the City University of New York \\
              PhD Program in Biochemistry, Graduate Center of the City University of New York \\
              \email{egallicchio@brooklyn.cuny.edu}
           }

\date{}

\maketitle

\begin{abstract}

We report the results of our participation in the SAMPL8 GDCC Blind Challenge for host-guest binding affinity predictions. Absolute binding affinity prediction is of central importance to the biophysics of molecular association and pharmaceutical discovery. The blinded SAMPL series have provided an important forum for assessing the reliability of binding free energy methods in an objective way. In this blinded challenge, we employed two binding free energy methods, the newly developed alchemical transfer method (ATM) and the well established potential of mean force (PMF) physical pathway method, using the same setup and force field model. The calculated binding free energies from the two methods are in excellent quantitative agreement. Importantly, the results from the two methods were also found to agree well with the experimental binding affinities released subsequently, with an $R^2$ of 0.89 (ATM) and 0.83 (PMF). Given that the two free energy methods are based on entirely different thermodynamic pathways, the close agreement between the results from the two methods and their general agreement with the experimental binding free energies are a testament to the the high quality achieved by theory and methods. The study provides further validation of the novel ATM binding free energy estimation protocol and it paves the way to to further extensions of the method to more complex systems.

\end{abstract}

\section{\label{sec:intro}Introduction}

The Statistical Assessment of Modeling of Proteins and Ligands (SAMPL) series of community challenges\cite{geballe2010sampl2,mobley2014blind,amezcua2021sampl7} have been organized to validate computational methods of molecular solvation and binding in an unbiased way. SAMPL participants are asked to quantitatively predict experimental measurements that are publicly disclosed only after the predictions are submitted. The format of the challenges allows the robust assessment of computational methods and have significantly contributed to their advancement.\cite{mobley2017predicting} As computational models of small molecule binding to protein receptors increasingly emerge as important elements of structure-based drug discovery,\cite{Jorgensen2009,armacost2020novel} it is critical that the reliability of these models is independently assessed and validated. We have contributed to several editions of the SAMPL challenges to validate the ability of our computational models to accurately predict host-guest and protein-ligand binding affinities.\cite{Gallicchio2012a,Gallicchio2014octacid,GallicchioSAMPL4,deng2016large,pal2016SAMPL5}.

In this work, we apply two conceptually orthogonal yet equivalent binding free energy estimation methods, the Alchemical Transfer Method (ATM)\cite{wu2021alchemical} and the Potential of Mean Force (PMF)\cite{deng2018comparing} method, to the SAMPL8 GDCC challenge set\footnote{\tt github.com/\-samplchallenges/\-SAMPL8/\-tree/\-master/\-host\_guest/\-GDCC }. The modeled predictions are tested against each other, as well as with the blinded experimental binding free energies measured by the Gibb Group.\cite{suating2020proximal}\footnote{ \tt github.com/\-samplchallenges/\-SAMPL8/\-blob/\-master/\-host\_guest/\-Analysis/\-ExperimentalMeasurements/\-Final-Data-Table-031621-SAMPL8.docx}

In principle, computational models should yield equivalent binding free energy predictions as long as they are based on the same chemical model and physical description of inter-atomic interactions. By ensuring consistency between two independent computational estimates, we can achieve an increased level of confidence in the theoretical accuracy of the models and in the correctness of their implementation. Furthermore, by comparing the computational predictions to the experimental measurements in a blinded, unbiased fashion, we can assess the predictive capability that can be expected of the models in actual chemical applications.

While a variety of empirical methods are commonly used to model the binding affinities of molecular complexes,\cite{sledz2018protein,seidel2020applications} here we are concerned with methods based on physical models of inter-atomic interactions and a rigorous statistical mechanics theory of the free energy of molecular binding.\cite{Gilson:Given:Bush:McCammon:97,Gallicchio2011adv,cournia2020rigorous} Binding free energy methods are classified as physical or alchemical depending on the nature of the thermodynamic path employed to connect the unbound to the bound states of the molecular complex for computing the reversible work of binding.\cite{Gallicchio2021binding} Physical pathway methods define a physical path in coordinate space in which the reversible work for bringing the two molecules together is calculated. Conversely, alchemical methods connect the bound and unbound states by a series of artificial intermediate states in which the ligand is progressively decoupled from the solution environment and coupled to the receptor. 

In this work, we compare the results of the PMF method,\cite{deng2018comparing} a physical pathway method, to that of the ATM alchemical method\cite{wu2021alchemical} on identically prepared molecular systems. Because free energy is a thermodynamic state function, binding free energy estimates should be independent of the specific path employed, whether physical or alchemical. Obtaining statistically equivalent estimates of the binding free energies using these two very different thermodynamic paths constitutes a robust validation of both methods. The very recently developed ATM, in particular, benefits from the backing of the more established PMF method in this application. 

This paper is organized as follows. We first review the PMF and ATM methods, describe the host-guest systems included in the SAMPL8 GDCC challenge, and provide the system setup and simulation details of our free energy calculations. We then present the binding free energy estimates we obtained with the PMF and ATM approaches and compare them to each other and with the experimental measurements that were disclosed only after the predictions were submitted to the SAMPL8 organizers.
Overall, the work shows that the ATM and PMF methods provide consistent binding free energy estimates that, in conjunction with the force field model employed here, are in statistical agreement with experimental observations. 

\section{\label{sec:methods}Theory and Methods}

\subsection{The Potential of Mean Force Method}

The Potential of Mean Force method, hereon PMF, employed in this work is a physical binding pathway approach fully described in reference \citenum{deng2018comparing}. Here, we briefly summarize the statistical mechanics basis of the method. Implementation details specific to this work are described in the Computational Details section.

\begin{figure*}
    \begin{center}
     \includegraphics[scale = 0.45]{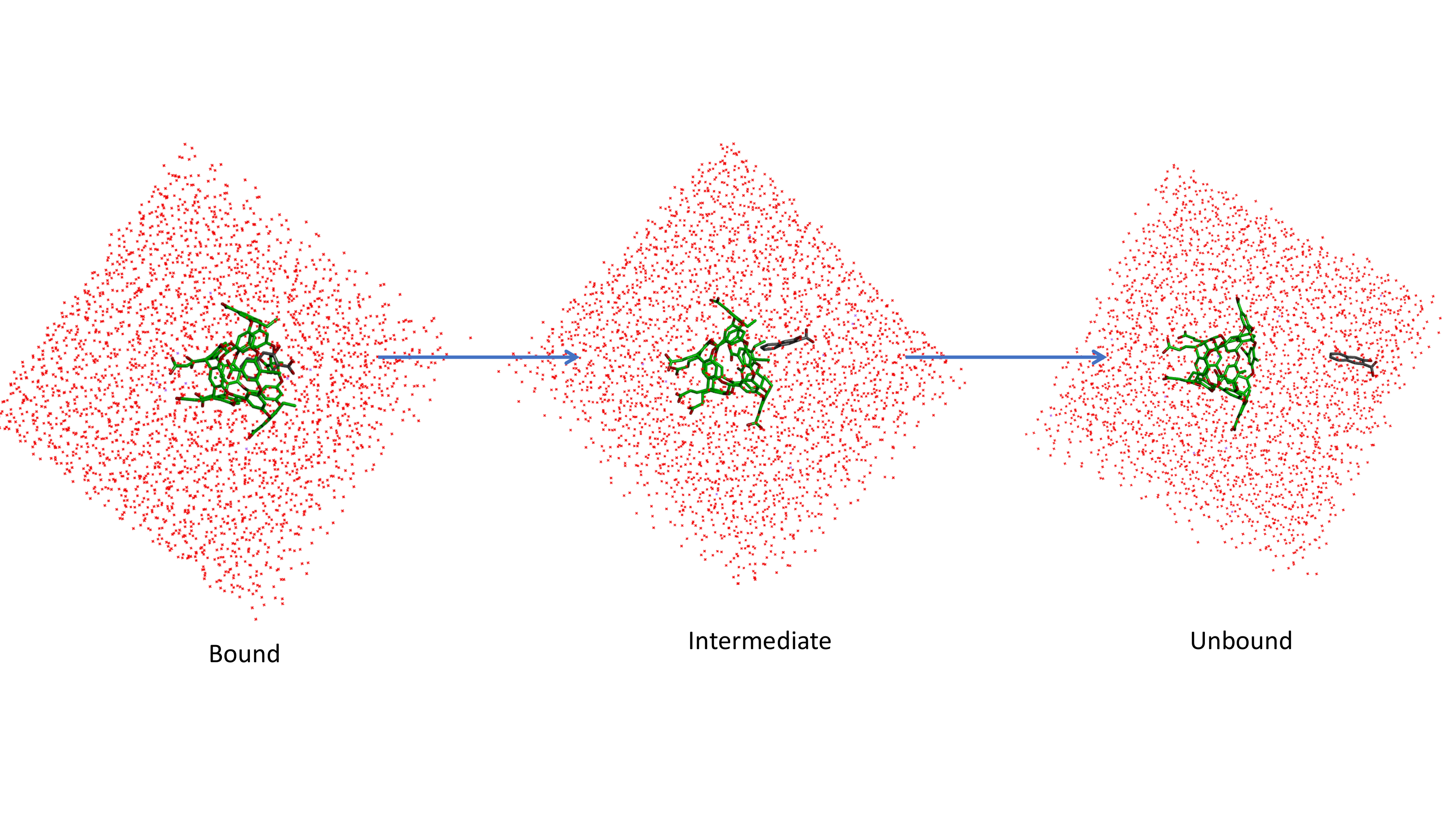}
    \end{center}
    \caption{Schematic of Potential of Mean Force (PMF) method. From left to right, the figure represents the physical pathway that the ligand undergoes from the bound to unbound state. Shown above is a sequence of 3 snapshots representing 3 of the 20 umbrella windows, where the ligand gets pulled at varying distances along the physical pathway away from the host (through the use of reference atoms assigned to both the ligand and host). The red dots represent the oxygen atoms of water molecules. The big bulky molecule represents the TEMOA host, while the small molecule represents the G1 guest.}
    \label{fig:PMF}
\end{figure*}

The PMF method estimates the standard free energy of binding as the sum of the free energy changes of the following processes: 
\begin{enumerate}
    \item The transfer of one ligand molecule from an ideal solution at the standard concentration $C^\circ = 1 M$ to a region in the solvent bulk of volume equal to the volume of the receptor binding site, followed by the imposition of harmonic restraints that keep the ligand in a chosen reference binding orientation. The free energy term corresponding to this process, denoted as $\Delta G^{\rm bulk}_{\rm restr}$, is evaluated analytically. 
    \item The transfer of the ligand molecule from the solvent bulk to the receptor binding site along a suitable physical pathway (see Computational Details). The free energy change along this pathway is described by a potential of mean force parameterized by the distance between two reference atoms of the ligand and the receptor (Figure \ref{fig:PMF}). The free energy change for this process, denoted by $w(r_{\rm min})-w(r^\ast)$, is given by the value at the minimum of the potential of mean force relative to the value in the bulk. 
    \item $\Delta G_{\rm vibr}$ is related to the ratio of the configurational partition functions of the ligand within the binding site of the receptor vs. when it is harmonically restrained at the bulk location $r^\ast$.
    \item The release of the harmonic restraints while the ligand is bound to the receptor. The free energy change for this process, denoted by $-\Delta G_{\rm restr}^{\rm bound}$, is evaluated by Bennett’s Acceptance Ratio method (BAR).
\end{enumerate}

Hence, the PMF estimate of the free energy of binding is given by
\begin{equation}
    \Delta G^\circ_b = \Delta G^{\rm bulk}_{\rm restr} + [ w(r_{\rm min})-w(r^\ast) ] + \Delta G_{\rm vibr} - \Delta G_{\rm restr}^{\rm bound}
    \label{eq:pmf-deltag}
\end{equation}
Additional computational details and parameters used in this work to implement the PMF calculations are described in the Computational Details section. 

\subsection{The Alchemical Transfer Method}

The Alchemical Transfer Method, hereon ATM, is a recently-developed method to compute the absolute binding free energy of molecular complexes. The method is fully described in reference \citenum{wu2021alchemical}. Here, we give only a brief overview of ATM, particularly focusing on the aspects specific to this work. Further implementation details are described in the Computational Details section.

Given the standard free energy of binding $\Delta G^\circ_b$, defined as the difference in free energy between the bound complex and the unbound components, $\Delta G^\circ_{\rm b} = \Delta G^\circ_{\rm site} + \Delta G^\ast_b $. ATM computes the excess component of the binding free energy, $\Delta G^\ast_b$, defined as the reversible work for transferring the ligand from a region of volume $V_{\rm site}$ in the solvent bulk to a region of the same volume in the receptor binding site.\cite{Gallicchio2011adv} The standard free energy of binding is given by the excess component plus the ideal component, $\Delta G^\circ_{\rm site} = -k_B T \ln C^\circ V_{\rm site}$, which corresponds to the free energy change of transferring one ligand molecule from an ideal solution at the standard concentration $C^\circ = 1 M$ to a region in the solvent bulk of volume that is equal to the volume of the receptor binding site, $V_{\rm site}$.\cite{Gilson:Given:Bush:McCammon:97} The concentration-dependent ideal term is computed analytically and the excess component is computed by ATM using numerical molecular simulations described in Computational Details and below.

\begin{figure*}
    \begin{center}
     \includegraphics[scale = 0.75]{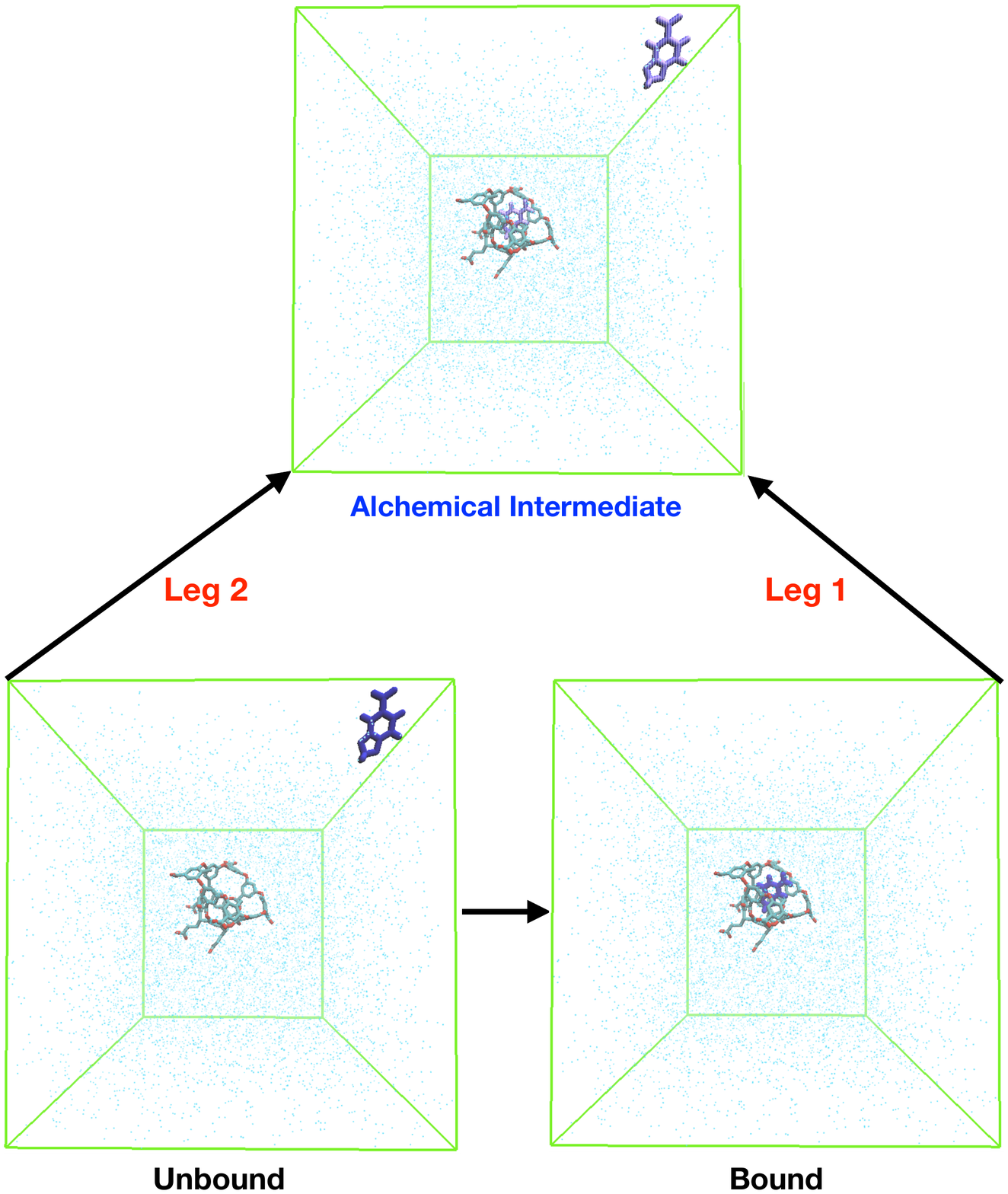}
    \end{center}
    \caption{The Alchemical Transfer Method (ATM) involves two simulation legs, which, in total, transfer the ligand from the solvent bulk to the binding site of the receptor. The two legs connect the bound and unbound end states through an alchemical intermediate that involves the ligand molecule interacting equally with both the receptor and the solvent bulk at half strength. Here, the receptor is the TEMOA host and the ligand is the G4 guest. The green box represents the solvent box with water molecules designated in blue. In the TEMOA structure, carbon atoms are represented in cyan and oxygen atoms in red.}
    \label{fig:ATM}
\end{figure*}

In ATM, the transfer of the ligand from the solvent bulk to the receptor binding site is carried out in two alchemical steps that connect the bound and unbound end states to one alchemical intermediate (Figure \ref{fig:ATM}), in which the ligand molecule interacts equally with both the receptor and the solvent bulk at half strength. The potential energy function of the alchemical intermediate is defined as
\begin{equation}
U_{1/2}(x_S, x_L) = \frac12 U(x_S, x_L) + \frac12 U(x_S, x_L + h) \, ,
\label{eq:pot_alchemical_hybrid_intermediate}
\end{equation}
where $x_S$ denotes the coordinates of the atoms of the receptor and of the solvent, $x_L$ denotes the coordinates of the atoms of the ligand while in the receptor binding site, and $h$ is the constant displacement vector that brings the atoms of the ligand from the receptor site to the solvent bulk site. In this scheme, $U(x_S, x_L)$ is the potential energy of the system when the ligand is in the binding site, $U(x_S, x_L + h)$ is the potential energy after translating the ligand rigidly into the solvent bulk, and $U_{1/2}(x_S, x_L)$ is the hybrid alchemical potential given by the average of the two. In the alchemical intermediate state, receptor atoms and solvent molecules interact with the ligand at half strength but at both ligand locations. Similarly, the force that ligand atoms interact with receptor atoms and solvent molecules at the intermediate state is an average of the forces exerted by the ligand at the two distinct locations. As discussed in reference \citenum{wu2021alchemical}, the ATM alchemical intermediate has an analogous role as the vacuum intermediate state in the conventional double-decoupling method,\cite{Gilson:Given:Bush:McCammon:97} but without fully dehydrating the ligand.  

The bound and unbound states of the complex are connected to the common intermediate by means of alchemical potentials of the form
\begin{equation}
U_{\lambda}(x)=U_{0}(x)+\lambda u_{\rm sc}[u(x)],
\label{eq:pert_pot}
\end{equation}
where $U_0(x)$ denotes the potential energy function of the initial state, which is either $U(x_S, x_L)$, corresponding to the bound complex in Leg 1 (Figure \ref{fig:ATM}), or $U(x_S, x_L + h)$, corresponding to Leg 2 (Figure \ref{fig:ATM}), $\lambda$ is a progress parameter that goes from $0$ to $1/2$,
\begin{equation}
u(x)= U_{1}(x)-U_{0}(x) \label{eq:sc-binding-energy}
\end{equation}
is the binding energy function.\cite{Gallicchio2010} In Equation \ref{eq:sc-binding-energy}, $U_1(x)$ denotes the potential energy function of the end state which is either $U(x_S, x_L + h)$, corresponding to the unbound complex in Leg 1 of Figure \ref{fig:ATM}, or $U(x_S, x_L)$, corresponding to the bound complex in Leg 2 (Figure \ref{fig:ATM}). Finally, 
\begin{equation}
  u_{\rm sc}(u)= u; \quad u \le u_c
\end{equation}
\begin{equation}
u_{\rm sc}(u)=(u_{\rm max} - u_c ) f_{\rm sc}\left[\frac{u-u_c}{u_{\rm max}-u_c}\right] + u_c; \quad u > u_c
\label{eq:soft-core-general}
\end{equation}
with
\begin{equation}
f_{\rm sc}(y) = \frac{z(y)^{a}-1}{z(y)^{a}+1} \label{eq:rat-sc} \, ,
\end{equation}
and
\begin{equation}
    z(y)=1+2 y/a + 2 (y/a)^2
\end{equation}
is a soft-core perturbation function that avoids singularities near the initial states of each leg ($\lambda = 0$). The parameters of the soft-core function, $u_{\rm max}$, $u_c$, and $a$ used in this work are listed in Computational Details.

The free energy change for each leg is obtained by multi-state thermodynamic reweighting\cite{Tan2012} using the perturbation energies $u_{\rm sc}[u(x)]$ collected during the molecular dynamics runs at various values of $\lambda$. As illustrated by the thermodynamic cycle in Figure \ref{fig:ATM}, the excess component of the binding free energy is obtained by the difference of the free energies of the two legs:
\begin{equation}
  \Delta G^\ast_b = \Delta G_2 - \Delta G_1 \, .
  \label{eq:cycle_dg}
\end{equation}

Because the end states of ATM are similar to that of the PMF method summarized above, the two methods compute the same free energy of binding. However, each employs a different thermodynamic path. The PMF method progressively displaces the ligand from the binding site to the bulk along a physical path, whereas ATM employs an unphysical alchemical path, in which the ligand is displaced directly from the binding site to the solvent bulk.

\subsection{SAMPL8 Systems}

The chemical structures of the two hosts and 5 guests molecules are shown in Fig. 3. Both the hosts TEETOA and TEMOA are octaacids that carry a net charge of -8 at the pH value of 11.5 used in the experiment. The five guests, with the exception of the protonated G2 (namely G2P), are carboxylate derivatives that are also negatively charged at the same pH. 
The computational calculations employed the initial host and guest structure files provided in the SAMPL8 dataset found at {\tt https://github.com\-/samplchallenges\-/SAMPL8\-/tree\-/master\-/host\_guest\-/GDCC}.

\begin{figure*}
    \centering
    \includegraphics[scale = 0.7]{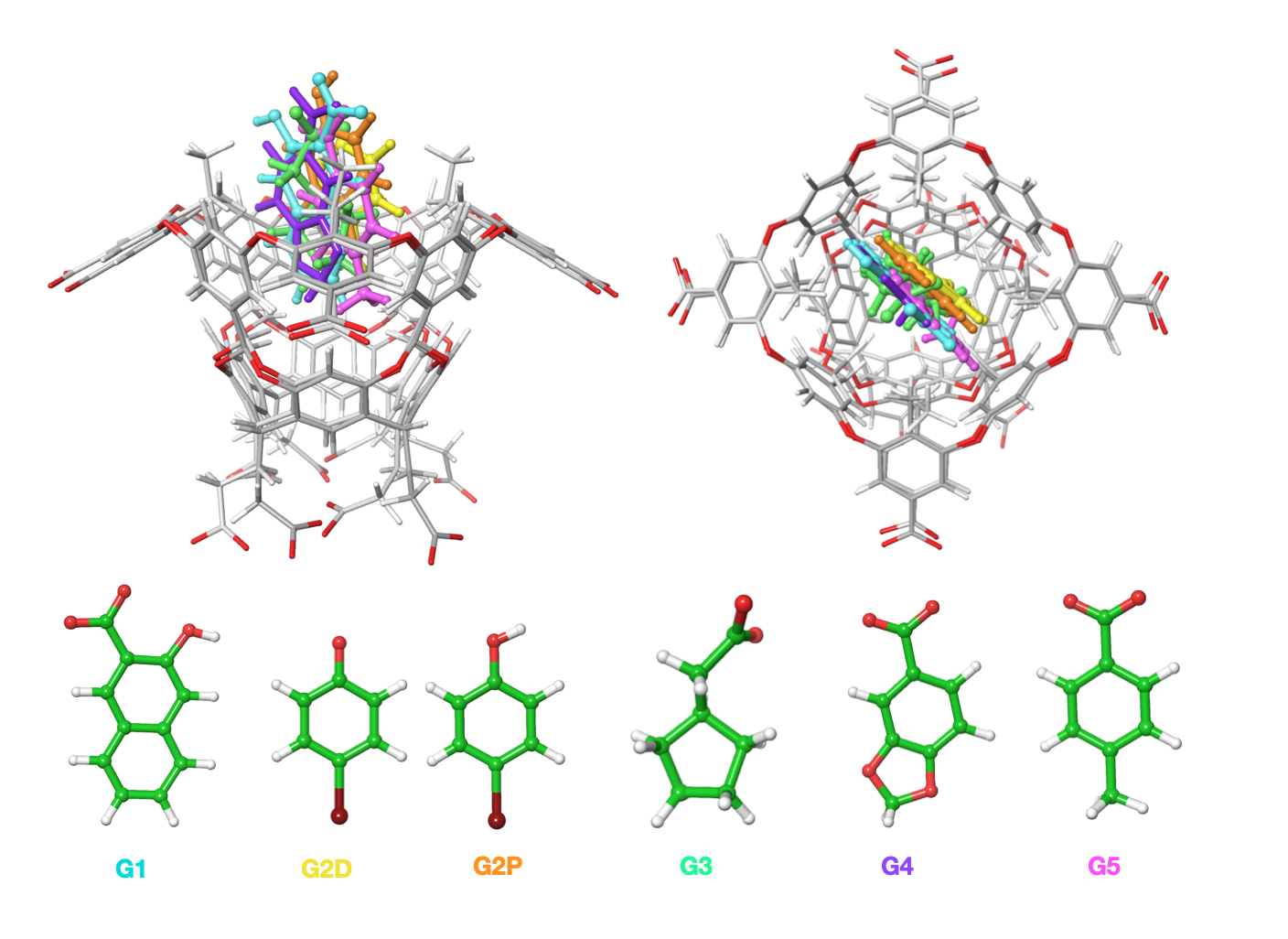}
    \caption{Superimposed benchmark systems in this study. The two hosts, tetramethyl octa acid (TEMOA) and tetraethyl octa acid (TEETOA), are shown in licorice representation, with light gray corresponding to TEETOA and dark gray to TEMOA. Both light and dark gray represent carbon atoms and red, oxygen atoms. The six guests that are bound to the hosts are shown in ball-and-stick (CPK) representation, for which the color of the structure corresponds to the label of the guest. G2D designates deprotonated G2 and G2P, protonated G2. Note that ball-and-stick representation undermines the aromaticity of the six-membered ring. For the guests, green corresponds to carbon atoms, red oxygen atoms, and white hydrogen atoms.}
    \label{fig:structures}
\end{figure*}

\subsection{Computational Details}

 The guests were manually docked to each host using Maestro (Schr\"{o}dinger, Inc.) to render a set of host-guest molecular complexes that were then used to derive forcefield parameters with AmberTools. The complexes were assigned GAFF2/AM1-BCC parameters and solvated in a water box with a 12 Angstrom solvent buffer and sodium counterions to balance the negative charge. The position and orientation of the host for each complex were restrained near the center of the box and along the diagonal with a flat-bottom harmonic potential of force constant 25.0 kcal/(mol \AA$^2$) and a tolerance of 1.5 \AA\ was set on the heavy atoms at the lower cup of the molecule (the first 40 atoms of the host as listed in the provided files). The systems were energy minimized and thermalized at 300 K prior to proceeding with the ATM and PMF calculations. 

\subsubsection{PMF setup}

The computation of the standard binding free energies using the PMF method involves the following steps:\cite{deng2018comparing} (1) applying a harmonic restraint on the three Euler angles of the guest in the bound state to restrain guest orientation; (2) applying a harmonic restraint on the polar and azimuthal angles in spherical coordinates to restrain the guest center along a fixed axis when it binds/unbinds; (3) reversibly extracting the guest from the binding pocket along the chosen axis until it reaches the bulk region; (4) release the restraints on the guest center and guest orientation, which allows the guest to occupy the standard volume and rotate freely in the bulk solvent. The standard binding free energy is then obtained by summing up the reversible work associated with each of the above steps using Eq.\ (\ref{eq:pmf-deltag}).

The position and orientation of the guest relative to the host was controlled using coordinate systems which consisted of 3 reference atoms of the host (P1, P2, and P3) and 3 reference atoms of the guest (L1, L2, and L3).\cite{Boresch:Karplus:2003}  For all the hosts, P1 was chosen to be the center of the bottom ring of each host and L1 the center of each guest molecule which lies approximately 4 Angstroms away from P1. The PMF was calculated along the P1-L1 distance using umbrella sampling with biasing potentials having a force constant of 1000 kJ/(mol nm$^2$). The three Euler angles and two polar and azimuthal angles were restrained using harmonic potentials with a force constant of 1,000 kJ/(mol rad$^2$) centered on the angles of the thermalized structures such that the guest is pulled straight out of the pocket of the host while minimizing collisions with the sidechains of the rim of the host. It is important to note that an unobstructed path is necessary for the guest's pull axis for the PMF method.

Equilibration (1.2 ns) and production (20 ns) umbrella sampling was then initiated over 20 umbrella windows to cover a distance of 4.0 to 18.0 Angstroms, i.e. from within the binding region to the bulk along the P1-L1 axis. WHAM analysis was used  to generate the PMF and the corresponding uncertainties by bootstrapping. The free energy of releasing the angular restraints in the bulk and in the bound state were computed using BAR as implemented in GROMACS.\cite{pronk2013gromacs}

\subsubsection{ATM Setup}

Each of the Cartesian components of the translation vector $h$ were set to approximately half of the longest diagonal of the simulation box to place the ligand near the corner of the solvent box farthest away from the host and its periodic images (Fig. \ref{fig:ATM}). Beginning at the bound state at $\lambda=0$, the systems were then progressively annealed to the symmetric alchemical intermediate at $\lambda = 1/2$ during a $250$ ps run using the ATM alchemical potential energy function for Leg 1 [Eq.\ (\ref{eq:pot_alchemical_hybrid_intermediate})]. This step yields a suitable initial configuration of the system without severe unfavorable repulsive interactions at either end states of the alchemical path so that molecular dynamics replica exchange alchemical simulation can be conducted for each leg as described below. 

In order to prevent large attractive interactions between opposite charges at small distances in nearly uncoupled states, polar hydrogen atoms with zero Lennard-Jones parameters were modified to $\sigma_{\rm LJ} = 0.1$ \AA\ and $\epsilon_{\rm LJ} = 10^{-4}$ kcal/mol. \cite{khuttan2021single} We established that the change in potential energy of the system in the unbound, bound, and symmetric intermediate states due to this modification of the Lennard-Jones parameters is below single floating point precision. Alchemical MD calculations were conducted with the OpenMM 7.3\cite{eastman2017openmm} MD engine and the SDM integrator plugin ({\tt github.com/\-Gallicchio-Lab/\-openmm\_sdm\_plugin.git}) using the OpenCL platform. In order to maintain the temperature at 300 K, a Langevin thermostat with a time constant of 2 ps was implemented. 
For each ATM leg, Hamiltonian Replica Exchange in $\lambda$ space was conducted every 5 ps with the ASyncRE software \cite{gallicchio2015asynchronous} that is customized for OpenMM and SDM ({\tt github.com/\-Gallicchio-Lab/\-async\_re-openmm.git}). Each leg employed 11 $\lambda$ states uniformily distributed between $\lambda=0$ and $\lambda=1/2$. All ATM calculations employed the soft-core perturbation energy with parameters $u_{\rm max} = 300$ kcal/mol, $u_c = 100$ kcal/mol, and $a=1/16$. A flat-bottom harmonic potential between the centers of mass of the host and the guest with a force constant of 25 kcal/mol $\AA^2$ was applied for a distance greater than $4.5 \AA$ to define the binding site region ($V_{\rm site}$).  The concentration-dependent term, $\Delta G^\circ_{\rm site} = -k_B T \ln C^\circ V_{\rm site} = 0.87$, which corresponds to 300 K temperature and the volume $V_{\rm site}$ of a sphere with a radius of $4.5  \AA$, was added to yield the final free energy estimate. Perturbation energy samples and trajectory frames were collected every 5 ps. Every replica was simulated for a minimum of 10 ns. For ATM, UWHAM was used to compute binding free energies and the corresponding uncertainties with the first 5 ns of the trajectory discarded. 

\subsubsection{Free Energy of Binding for Ligands in Multiple Protonation States}

When multiple chemical species contribute to binding, we use the free energy combination formula\cite{Gallicchio2011adv}
\begin{equation}
\Delta G_{b}^{\circ}=-kT\ln\sum_{i}P_{0}(i)e^{-\beta\Delta G_{b}^{\circ}(i)},
\label{eq:DG0sum}
\end{equation}
where $\Delta G_{b}^{\circ}(i)$ is the standard binding free energy for species $i$ and $P_0(i)$ is the population of that species in the unbound state. In the case of an acid/base equilibrium with acidity constant
\begin{equation}
    K_a = \frac{[A^{-}] [H^{+}]}{[HA]} = \frac{[A^{-}] }{[HA]} 10^{-pH} = \alpha 10^{-pH},
\end{equation}
where $[\ldots]$ are concentration in molar units,  
\begin{equation}
    \alpha = 10^{pH - pKa}, 
\end{equation}
is the concentration ratio of the deprotonated and protonated forms, the population fraction of the deprotonated species is
\begin{equation}
    P_0(A^{-}) =  \frac{[A^{-}]}{[HA]+[A^{-}]} = \frac{\alpha}{1+\alpha}
\end{equation}
and the population fraction of the protonated species is
\begin{equation}
    P_0(HA) =  \frac{[HA]}{[HA]+[A^{-}]} = 1 - P_0(A^{-}) = \frac{1}{1+\alpha} .
\end{equation}
The populations of each protonation state of the ligands and the corresponding standard binding free energies $\Delta G_b^\circ(A^{-})$ and $\Delta G_b^\circ(HA)$ are combined using Eq.\ (\ref{eq:DG0sum}) to obtain an estimate of the observed free energy of binding.

This strategy was employed for the guest G2, 4-bromophenol, which exists in two protonation states. A pH of 11.5, as indicated in the SAMPL8 GitHub site, and a pKa of 9.17 ({\tt pubchem.ncbi.nlm.nih.gov/\-compound/\-4-bromophenol}) was used to calculate the concentrations of the protonation states and combine them with the calculated binding free energies to yield a binding free energy estimate for G2 (see Table \ref{tab:protonation-stats}).

\section{Results}

The results are presented as follows. 
Table \ref{tab:calcs_vs_expt} summarizes the absolute binding free energy predictions from ATM  and PMF submitted to the SAMPL8 organizers, compared to the experimental values which were disclosed to us only after submission. The results of the constituent calculations for each method that led to the binding free energy predictions are listed in Tables \ref{tab:ATM-simulation-results} and \ref{tab:PMF-simulation-results} for the ATM and PMF methods, respectively. These tables report the values of the free energy changes for each leg of the ATM calculations and the components of the PMF estimates, including those of the vibrational free energy and the restraint free energy that contribute to the overall PMF process. The free energy analysis for the protonated and deprotonated species implicated in the complexes of the G2 guest is illustrated in Table \ref{tab:protonation-stats}.

\subsection{Absolute Binding Free Energy Estimates by ATM and PMF}

\begin{table}
    \centering
    \caption{PMF and ATM standard binding free energy predictions compared to the  experimental values.
    }
    \begin{tabular}{lccc}
         Complex & Experiment$^a$ & ATM$^a$ & PMF$^a$ \\ \hline
    TEMOA-G1 & $-6.96 \pm 0.2$  & $-6.71 \pm 0.3$ & $-6.43 \pm 0.4$ \\
    TEMOA-G2 & $-8.41 \pm 0.1$  & $-9.90 \pm 0.8$ & $-9.37 \pm 0.8$ \\
    TEMOA-G3 & $-5.78 \pm 0.1$  & $-8.26 \pm 0.3$ & $-8.71 \pm 0.4$ \\
    TEMOA-G4 & $-7.72 \pm 0.1$  & $-8.63 \pm 0.3$ & $-8.79 \pm 0.6$ \\
    TEMOA-G5 & $-6.67 \pm 0.1$  & $-7.70 \pm 0.3$ & $-8.15 \pm 0.8$ \\
    TEETOA-G1 & $-4.49 \pm 0.2$ & $-1.07 \pm 0.3$ & $-1.38 \pm 0.8$ \\
    TEETOA-G2 & $-5.16 \pm 0.1$ & $-4.76 \pm 0.3$ & $-6.22 \pm 1.8$ \\
    TEETOA-G3 & NB              & $-1.65 \pm 0.3$ & $-1.42 \pm 0.8$ \\
    TEETOA-G4 & $-4.47 \pm 0.2$ & $-2.51 \pm 0.3$ & $-2.25 \pm 0.8$ \\
    TEETOA-G5 & $-3.32 \pm 0.1$ & $-2.82 \pm 0.3$ & $-3.36 \pm 1.9$ \\ \hline
    \end{tabular}
    \label{tab:calcs_vs_expt}
    \begin{flushleft}\small
    $^a$ In kcal/mol.
\end{flushleft}
\end{table}

\begin{table*}
    \centering
      \caption{Agreement metrics (root mean square error, RMSE, correlation coefficient of determination, $R^2$, slope of the linear regression, $m$, and Kendall rank order correlation coefficient, $\tau$) between the computed binding free energies and the experimental measurements.}
    \begin{tabular}{c|cccc}
              & RMSE  & $R^2$  & m    & $\tau$ \\ \hline
     ATM/PMF  & 0.60  & 0.99  & 1.05 & 1.00 \\ \hline
     Exp./ATM & 1.71  & 0.89  & 1.65 & 0.69$^{a}$ \\ \hline
     Exp./PMF & 1.79  & 0.83  & 1.50 & 0.69$^{a}$ \\ \hline
    \end{tabular}
    \begin{flushleft}\small
    $^a$ TEETOA-G3, a non-binder experimentally, was included in the $\tau$ calculation as the weakest complex.
    \end{flushleft}
    \label{tab:metrics}
\end{table*}

The binding free energy estimates obtained from the two complementary computational methods, ATM and PMF, are in very good agreement with an $R^2$ value of 0.965 and an RMSE value of 0.989(?) kcal/mol. In addition, the ranking of the binding free energies of the complexes between the ATM and PMF datasets is in perfect agreement. Both methods consistently estimated the complex with the most favorable binding free energy to be TEMOA-G2, with a free energy value of -9.90 kcal/mol predicted by ATM and -9.37 kcal/mol by PMF. The least favorable binding free energy was predicted for the complex TEETOA-G1 by both methods, -1.07 kcal/mol by ATM and -1.38 kcal/mol by PMF. Both methods predict that all of the guests bind TEMOA more favorably than TEETOA.

All of the carboxylic acid guests were modeled as ionic. We modeled both protonation states of the G2 guest (Tables \ref{tab:ATM-simulation-results} and \ref{tab:PMF-simulation-results}) and combined the corresponding binding free energies using the experimental pKa of the guest (Table \ref{tab:protonation-stats}). With a discrepancy of 2.77 kcal/mol, the deprotonated G2 molecule (hereon G2D) yielded the  most divergent binding free energy estimate  between the ATM and PMF datasets. Nevertheless, since this protonation state is found to contribute little to binding 
(Table \ref{tab:protonation-stats}), the observed discrepancy did not affect significantly the correspondence between the two sets of SAMPL8 binding free energy predictions.

The molecular dynamics trajectories consistently yielded the expected binding mode of the guests to the TEMOA and TEETOA hosts. The polar/ionic end of the guests is oriented towards the water solvent while the more non-polar end of the molecule is inserted into the binding cavity of the hosts (Figure \ref{fig:structures}). In the complexes, the ethyl sidechains of the TEETOA host point outward extending further the host binding cavity and the surface of contact between the guests and the hosts. In the apo state, however, the ethyl sidechains are observed to be mostly folded into the TEETOA cavity (not shown). We hypothesize that the conformational reorganization of TEETOA, the lack of favorable water expulsion,  and the poorer hydration of the bound guests are responsible for the weaker binding capacity of TEETOA relative to TEMOA. We intend to investigate further these aspects of the binding mechanism in future work. 

ATM and PMF both predict that G2D is one of the weakest binders for TEMOA  and TEETOA (Tables \ref{tab:ATM-simulation-results} and \ref{tab:PMF-simulation-results}). G2D is expected to be frustrated in the bound state because the bromine atom prefers to be in the cavity of the host, whereas the oxide group strongly prefers to remain hydrated (Figure \ref{fig:structures}). The side chains of both hosts prevent the hydration of the negative oxygen atom. This steric hindrance is especially evident in TEETOA, which possesses four ethyl groups on its outer ring. Due to its poor binding affinity, the deprotonated G2D is not predicted to contribute significantly to binding despite its higher concentration in solution at the experimental pH. Conversely, due to its smaller desolvation penalty, both the PMF and ATM methods indicate that protonated  G2 (hereon G2P) is the strongest binder in the set for both TEMOA and TEETOA (Tables \ref{tab:ATM-simulation-results} and \ref{tab:PMF-simulation-results}). G2P is in fact predicted to be the dominant species for binding even after factoring in the protonation penalty at the experimental pH of 11.5. 

The ATM free energy components $\Delta G_1$ and $\Delta G_2$ for each leg of the ionic hosts (Table \ref{tab:ATM-simulation-results}), being in the 40 to 50 kcal/mol range, are significantly larger in magnitude than the resulting binding free energies. These free energies correspond to the reversible work to reach the alchemical intermediate state in which the guest interacts with both the receptor and the solvent bulk intermediates. The high free energy of the alchemical intermediate relative to the bound and solvated states suggests that the ionic group can not be properly accommodated to simultaneously interact effectively with both environments. This hypothesis is confirmed by the much smaller ATM leg free energies for the neutral G2P guest. While large, the ATM leg free energies of the ionic guests are expected to be significantly smaller than those that would have obtained in a double-decoupling calculation\cite{deng2018comparing} that would involve displacing the guests into vacuum where hydration interactions are completely removed. The statistical uncertainties of the ATM binding free energy estimates, generally around $1/3$ of a kcal/mol, are relatively small. 

While still moderate, the PMF binding free energy estimates (Table \ref{tab:PMF-simulation-results}) come with somewhat larger uncertainties than ATM. The source of uncertainties is approximately equally split between the reversible work of releasing the restraints (2nd column) and work of ligand extraction (3rd column). However, in some cases (TEETOA-G2 and TEETOA-G5) the uncertainty of the work of extraction is particularly large and probably indicative of sampling bottlenecks at intermediate stages of the extraction process for this host.

\begin{table*}
\centering
\caption{ATM absolute binding free energy estimates for the TEMOA and TEETOA complexes.}
\label{tab:ATM-simulation-results}
\begin{tabular}{lcccc}
Complex    & $\Delta G_1$$^{a}$ & $\Delta G_2$$^{a}$ & $\Delta G^\circ_{\rm site}$$^{a}$ & $\Delta G^\circ_b$$^{a}$ \\ \hline 
TEMOA-G1   & $ 53.27 \pm 0.21 $ & $ 45.69 \pm 0.21 $ & $ 0.87 $ & $ -6.71  \pm 0.30 $  \\ 
TEMOA-G2D  & $ 42.37 \pm 0.18 $ & $ 35.48 \pm 0.21 $ & $ 0.87 $ & $ -6.02  \pm 0.28 $  \\ 
TEMOA-G2P  & $ 22.57 \pm 0.27 $ & $  8.60 \pm 0.78 $ & $ 0.87 $ & $ -13.10 \pm 0.83 $  \\
TEMOA-G3   & $ 56.42 \pm 0.18 $ & $ 47.29 \pm 0.18 $ & $ 0.87 $ & $ -8.26  \pm 0.25 $ \\ 
TEMOA-G4   & $ 53.13 \pm 0.24 $ & $ 43.63 \pm 0.18 $ & $ 0.87 $ & $ -8.63  \pm 0.30 $ \\ 
TEMOA-G5   & $ 53.49 \pm 0.24 $ & $ 44.92 \pm 0.18 $ & $ 0.87 $ & $ -7.70  \pm 0.30 $ \\ \hline
TEETOA-G1  & $ 51.65 \pm 0.27 $ & $ 49.71 \pm 0.21 $ & $ 0.87 $ & $ -1.07  \pm 0.34 $ \\ 
TEETOA-G2D & $ 42.26 \pm 0.24 $ & $ 39.83 \pm 0.27 $ & $ 0.87 $ & $ -1.57  \pm 0.36 $ \\ 
TEETOA-G2P & $ 22.31 \pm 0.24 $ & $ 13.48 \pm 0.15 $ & $ 0.87 $ & $ -7.95  \pm 0.28 $  \\ 
TEETOA-G3  & $ 55.31 \pm 0.24 $ & $ 52.79 \pm 0.18 $ & $ 0.87 $ & $ -1.65  \pm 0.30 $ \\ 
TEETOA-G4  & $ 52.28 \pm 0.24 $ & $ 48.90 \pm 0.18 $ & $ 0.87 $ & $ -2.51  \pm 0.30 $  \\ 
TEETOA-G5  & $ 53.58 \pm 0.21 $ & $ 49.89 \pm 0.18 $ & $ 0.87 $ & $ -2.82  \pm 0.28 $ \\ \hline
\end{tabular}
\begin{flushleft}\small
    $^a$ In kcal/mol.
\end{flushleft}
\end{table*}

\makesavenoteenv{tabular}
\begin{table*}
\centering
\caption{PMF absolute free energy estimates for TEMOA and TEETOA complexes.}
\label{tab:PMF-simulation-results}
\begin{tabular}{lcccccc}
Complex & $-\Delta G_{\rm restr}^{\rm bound}$$^{a}$ & $[w(r_{\rm min}) - w(r^\ast)]$$^{a}$ & $\Delta G_{\rm vibr}$$^{a}$ & $\Delta G_{\rm restr}^{\rm bulk}$$^{a}$ &  $\Delta G^\circ_b$$^{a}$ \\ \hline 
TEMOA-G1    &  $ -4.09 \pm 0.23 $ & $ -12.27 \pm 0.36 $   & $ 0.24 $    & $ 9.69 $  & $ -6.43 \pm 0.43 $  \\ 
TEMOA-G2D   &  $ -2.05 \pm 0.33 $ & $ -11.01 \pm 0.18 $   & $ 0.12 $    & $ 9.69 $  & $ -3.25 \pm 0.38 $  \\ 
TEMOA-G2P   &  $ -5.31 \pm 0.78 $ & $ -17.12 \pm 0.21 $   & $ 0.17 $    & $ 9.69 $  & $ -12.57 \pm 0.81 $  \\
TEMOA-G3    &  $ -5.61 \pm 0.30 $ & $ -12.83 \pm 0.30 $   & $ 0.04 $    & $ 9.69 $  & $ -8.71 \pm 0.42 $  \\ 
TEMOA-G4$^{b}$   &  $ -5.00 \pm 0.47 $ & $ -13.72 \pm 0.36 $   & $ 0.24 $    & $ 9.69 $  & $ -8.79 \pm 0.59$  \\ 
TEMOA-G5    &  $ -5.36 \pm 0.81 $ & $ -12.74 \pm 0.15 $   & $ 0.26 $    & $ 9.69 $  & $ -8.15 \pm 0.82 $  \\ \hline
TEETOA-G1   &  $ -3.76 \pm 0.60 $ & $ -7.60 \pm 0.54 $    & $  0.28 $   & $ 9.69 $  & $ -1.38 \pm 0.81$   \\ 
TEETOA-G2D  &  $ -5.50 \pm 0.84 $ & $ -5.25 \pm 2.73 $    & $  0.20 $   & $ 9.69 $  & $ -0.86 \pm 2.86 $   \\ 
TEETOA-G2P  &  $ -4.85 \pm 0.57 $ & $ -14.51 \pm 1.68 $   & $  0.25 $   & $ 9.69 $  & $ -9.42 \pm 1.77 $   \\ 
TEETOA-G3   &  $ -3.70 \pm 0.24 $ & $ -7.36 \pm 0.81 $    & $ -0.05 $   & $ 9.69 $  & $ -1.42 \pm 0.84 $   \\ 
TEETOA-G4   &  $ -3.77 \pm 0.12 $ & $ -8.39 \pm 0.75 $    & $  0.22 $   & $ 9.69 $  & $ -2.25 \pm 0.76 $ \\ 
TEETOA-G5   &  $ -4.47 \pm 0.06 $ & $ -8.81 \pm 1.89 $    & $  0.23 $   & $ 9.69 $  & $ -3.36 \pm 1.89 $   \\ \hline
\end{tabular}
\begin{flushleft}\small
$^{a}$ In kcal/mol.
\end{flushleft}
\end{table*}





\subsection{Calculated Free Energy Estimates Relative to Experimental Measurements}

The two computational methods employed in this work reproduced the experimental binding free energy estimates relatively well, particularly more so for the TEMOA host than for the TEETOA host (Table \ref{tab:calcs_vs_expt}). Both methods correctly predict TEMOA-G2 as the highest affinity complex in the set with good quantitative accuracy in the binding free energy predictions ($-8.41$ kcal/mol experimentally compared to calculated $-9.90$ and $-9.37$ kcal/mol from ATM and PMF, respectively). Concomitantly, both methods correctly predict relatively weak absolute binding free energies of -1.65 kcal/mol and -1.42 kcal/mol, respectively, for TEETOA-G3 which is an experimental non-binder. Excluding TEETOA-G3, the least favorable binding affinity measurement was obtained for TEETOA-G5, which is correctly scored as one of the weakest complex by both computational methods. Overall, despite the the narrow range of the moderate binding free energies, the computational rankings based on the binding free energies are in good agreement with the experimental rankings with a Kendall rank-order correlation coefficient of 0.69. (Table \ref{tab:metrics})

\begin{figure*}
    \centering
    \includegraphics[scale=0.6]{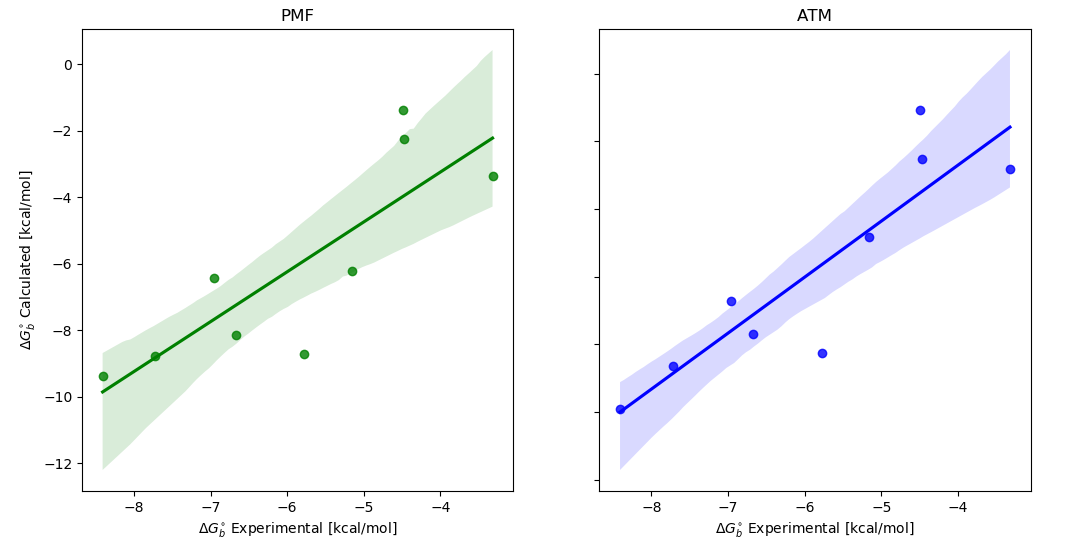}
    \caption{Linear regression of combined TEMOA and TEETOA predictions with ATM and PMF.}
    \label{fig:calcs_vs_expt}
\end{figure*}

As illustrated in Figure \ref{fig:calcs_vs_expt} the calculated binding free energies are highly correlated to the experimental values with Pearson $R^2$ correlation coefficients of 89\% and 83\% for ATM and PMF, respectively (Table \ref{tab:metrics}). The calculations are also in reasonable quantitative agreement with the experimental measurements with RMSE deviations of $1.71$ kcal/mol for ATM and $1.79$ kcal/mol for PMF. Interestingly, the computational models tend to overestimate the binding affinity of the TEMOA complexes and to underestimate those of the complexes with TEETOA.  The largest deviation occurs for TEETOA-G1 which has a moderate observed binding free energy of $-4.47$ kcal/mol, which is underestimated by the computational predictions by around $-1$ kcal/mol. A large deviation, but in the opposite direction, is also observed for TEMOA-G3 ($-5.78$ kcal/mol experimentally compared to $-8.26$ and $-8.71$ kcal/mol computationally) (Table v\ref{tab:calcs_vs_expt}). A poor prediction for this complex was expected based on previous efforts with the GAFF/AM1-BCC force field with TIP3P solvation used here.\cite{rizzi2018overview}

In summary, the blinded predictions reported here were scored as among the best of the SAMPL8 GDCC challenge and second only to those obtained with the more accurate AMOEBA force field\cite{shi2021amoeba} ({\tt github.com/\-samplchallenges/\-SAMPL8/\-blob/\-master/\-host\_guest/\-Analysis/\-Ranked\_Accuracy}).

\begin{table*}
    \centering
        \caption{Binding free energy contributions of the protonated and deprotonated G2 complexes to the ATM and PMF binding free estimates.}
    \begin{tabular}{lcccc}
    \hline
        & TEMOA-G2/ATM & TEMOA-G2/PMF & TEETOA-G2/ATM & TEETOA-G2/PMF  \\ \hline
      $\Delta G_{b}^\circ$(HA)$^{a}$   & $-13.10 \pm 0.83$ & $-12.57 \pm 0.81$ & $-7.95 \pm 0.28$ & $-9.42 \pm 1.77$ \\
      
      $P_0({\rm HA})$ & $4.66 \times 10^{-3}$ & $4.66 \times 10^{-3}$ & $4.66 \times 10^{-3}$ & $4.66 \times 10^{-3}$ \\
      
      \(\displaystyle P_{0}(HA)e^{-\beta\Delta G_{b}^{\circ}(HA)}\) & $1.65 \times 10^{7}$ & $6.77 \times 10^{7}$ & $2.92 \times 10^{3}$ & $3.42 \times 10^{4}$ \\ \hline
      
      $\Delta G_{b}^\circ({\rm A}^-)$$^{a}$ & $-6.02 \pm 0.28 $ & $-3.25 \pm 0.38$ & $-1.57 \pm 0.36$ & $-0.86 \pm 2.86$ \\
      
      $P_0({\rm A}^-)$ & $0.995$ & $0.995$ & $0.995$ & $0.995$ \\
      
      \(\displaystyle P_{0}(A^-)e^{-\beta\Delta G_{b}^{\circ}(A^-)}\) & $2.43 \times 10^{4}$ & $232$ & $13.6$ & $4.22$ \\ \hline
      
      $\Delta G_{b}^\circ$$^{a}$ & $-9.90 \pm 0.83$ & $-9.37 \pm 0.81$ & $-4.76 \pm 0.28$ & $-6.22 \pm 1.8$ \\ \hline 
    \end{tabular}
    \label{tab:protonation-stats}
    \begin{flushleft}\small
    $^a$ In kcal/mol.
\end{flushleft}
\end{table*}

\section{Discussion and Conclusions}

In this study, we employed two independent binding free energy approaches, the newly developed alchemical transfer method (ATM)\cite{khuttan2021single,wu2021alchemical} and the well established PMF physical pathway method\cite{deng2018comparing} to blindly predict the absolute binding affinities of the host-guest systems as part of the SAMPL8 GDCC blind challenge. The SAMPL series of community challenges has consistently yielded high-quality datasets to test computational models of binding,\cite{geballe2010sampl2,mobley2014blind,amezcua2021sampl7,GallicchioSAMPL4,deng2016large,pal2016SAMPL5} 
and we decided to use it here to stringently validate the ATM and PMF methods in an unbiased fashion. 

Despite their radical differences in spirit and in practice, we find that the calculated binding affinities from the two methods are in remarkable quantitative agreement with an RMSE of only 0.6 kcal/mol and an $R^2$ of $99$\%. This level of agreement, well within statistical fluctuations, gives high confidence in the theoretical foundations and in the correctness of implementation of each approach. The level of consistency of the computational methods also adds confidence that their predictions are unbiased and primarily reflective of the force field model. 

We find that the standard GAFF/AM1-BCC/TIP3P model employed here tends to overestimate the binding free energies of strongly bound complexes while it tends to understimate those of more weakly bound complexes, as also indicated by the larger than one slope of the linear regressions (Tables \ref{tab:calcs_vs_expt}, \ref{tab:metrics}). While it may be a result, in this case, of specific aspects of the TEMOA and TEETOA hosts, this trend has been generally observed with this force field combination.\cite{rizzi2018overview} The more accurate AMOEBA force field\cite{shi2021amoeba} appears to correctly predict these trends ({\tt github.com/\-samplchallenges/\-SAMPL8/\-blob/\-master/\-host\_guest/\-Analysis/\-Ranked\_Accuracy}).

The stringent blinded test conducted in this work is a further validation of the ATM binding free energy method that we have recently proposed.\cite{wu2021alchemical} ATM, implemented on top of the versatile OpenMM molecular dynamics engine,\cite{eastman2017openmm} promises to provide an accurate and streamlined route to absolute\cite{wu2021alchemical} and relative binding free calculations.\cite{Azimi2021RBFE} While alchemical, ATM, similar to the PMF pathway method,\cite{deng2018comparing} makes use of a single simulation system, and it avoids problematic vacuum intermediates and the splitting of the alchemical path into electrostatic and non-electrostatic transformations. ATM also does not require soft-core pair potentials and modifications of energy routines, and can be easily implemented as a controlling routine on top of existing force routines of MD engines.

In summary, this work provides a rare blinded and stringent test of binding free energy models. It shows that the application of sound statistical mechanics theories of binding and careful modeling of chemical systems can lead to reliable predictions limited only by the quality of the force field model.  

\section{Acknowledgements}

We acknowledge support from the National Science Foundation (NSF
CAREER 1750511 to E.G.). Molecular simulations were conducted on the
Comet and Expanse GPU clusters at the San Diego Supercomputing Center supported by NSF XSEDE award TG-MCB150001. We appreciate the National Institutes of Health for its support of the SAMPL project via R01GM124270 to David L. Mobley.


\begin{thebibliography}{10}

\bibitem{geballe2010sampl2}
Matthew~T Geballe, A~Geoffrey Skillman, Anthony Nicholls, J~Peter Guthrie, and
  Peter~J Taylor.
\newblock The {SAMPL2} blind prediction challenge: introduction and overview.
\newblock {\em J. Comp. Aided Mol. Des.}, 24(4):259--279, 2010.

\bibitem{mobley2014blind}
David~L Mobley, Shuai Liu, Nathan~M Lim, Karisa~L Wymer, Alexander~L Perryman,
  Stefano Forli, Nanjie Deng, Justin Su, Kim Branson, and Arthur~J Olson.
\newblock Blind prediction of hiv integrase binding from the {SAMPL4}
  challenge.
\newblock {\em J. Comp. Aided Mol. Des.}, pages 1--19, 2014.

\bibitem{amezcua2021sampl7}
Martin Amezcua, L{\'e}a El~Khoury, and David~L Mobley.
\newblock {SAMPL7} host--guest challenge overview: assessing the reliability of
  polarizable and non-polarizable methods for binding free energy calculations.
\newblock {\em J. Comp.-Aid. Mol. Des.}, 35(1):1--35, 2021.

\bibitem{mobley2017predicting}
David~L Mobley and Michael~K Gilson.
\newblock Predicting binding free energies: frontiers and benchmarks.
\newblock {\em Ann. Rev. Bioph.}, 46:531--558, 2017.

\bibitem{Jorgensen2009}
William~L Jorgensen.
\newblock Efficient drug lead discovery and optimization.
\newblock {\em Acc Chem Res}, 42:724--733, 2009.

\bibitem{armacost2020novel}
Kira~A Armacost, Sereina Riniker, and Zoe Cournia.
\newblock Novel directions in free energy methods and applications, 2020.

\bibitem{Gallicchio2012a}
E.~Gallicchio and R.~M. Levy.
\newblock Prediction of {SAMPL3} host-guest affinities with the binding energy
  distribution analysis method ({BEDAM}).
\newblock {\em J. Comp. Aided Mol. Design.}, 25:505--516, 2012.

\bibitem{Gallicchio2014octacid}
Emilio Gallicchio, Haoyuan Chen, He~Chen, Michael Fitzgerald, Yang Gao, Peng
  He, Malathi Kalyanikar, Chuan Kao, Beidi Lu, Yijie Niu, Manasi Pethe, Jie
  Zhu, and Ronald~M Levy.
\newblock {BEDAM} binding free energy predictions for the {SAMPL4} octa-acid
  host challenge.
\newblock {\em J. Comp. Aided Mol. Des.}, 29:315--325, 2015.

\bibitem{GallicchioSAMPL4}
Emilio Gallicchio, Nanjie Deng, Peng He, Alexander~L. Perryman, Daniel~N.
  Santiago, Stefano Forli, Arthur~J. Olson, and Ronald~M. Levy.
\newblock Virtual screening of integrase inhibitors by large scale binding free
  energy calculations: the {SAMPL4} challenge.
\newblock {\em J. Comp. Aided Mol. Des.}, 28:475--490, 2014.

\bibitem{deng2016large}
Nanjie Deng, William~F Flynn, Junchao Xia, RSK Vijayan, Baofeng Zhang, Peng He,
  Ahmet Mentes, Emilio Gallicchio, and Ronald~M Levy.
\newblock Large scale free energy calculations for blind predictions of
  protein--ligand binding: the d3r grand challenge 2015.
\newblock {\em J. Comp.-Aided Mol. Des.}, 30(9):743--751, 2016.

\bibitem{pal2016SAMPL5}
Rajat~Kumar Pal, Kamran Haider, Divya Kaur, William Flynn, Junchao Xia,
  Ronald~M. Levy, Tetiana Taran, Lauren Wickstrom, Tom Kurtzman, and Emilio
  Gallicchio.
\newblock A combined treatment of hydration and dynamical effects for the
  modeling of host-guest binding thermodynamics: The {SAMPL5} blinded
  challenge.
\newblock {\em J. Comp. Aided Mol. Des.}, 31:29--44, 2016.

\bibitem{wu2021alchemical}
Joe~Z Wu, Solmaz Azimi, Sheenam Khuttan, Nanjie Deng, and Emilio Gallicchio.
\newblock Alchemical transfer approach to absolute binding free energy
  estimation.
\newblock {\em J. Chem. Theory Comput.}, 17:3309, 2021.

\bibitem{deng2018comparing}
Nanjie Deng, Di~Cui, Bin~W Zhang, Junchao Xia, Jeffrey Cruz, and Ronald Levy.
\newblock Comparing alchemical and physical pathway methods for computing the
  absolute binding free energy of charged ligands.
\newblock {\em Phys. Chem. Chem. Phys.}, 20(25):17081--17092, 2018.

\bibitem{suating2020proximal}
Paolo Suating, Thong~T Nguyen, Nicholas~E Ernst, Yang Wang, Jacobs~H Jordan,
  Corinne~LD Gibb, Henry~S Ashbaugh, and Bruce~C Gibb.
\newblock Proximal charge effects on guest binding to a non-polar pocket.
\newblock {\em Chemical Science}, 11(14):3656--3663, 2020.

\bibitem{sledz2018protein}
Pawe{\l} {\'S}led{\'z} and Amedeo Caflisch.
\newblock Protein structure-based drug design: from docking to molecular
  dynamics.
\newblock {\em Curr. Op. Struct. Biol.}, 48:93--102, 2018.

\bibitem{seidel2020applications}
Thomas Seidel, Oliver Wieder, Arthur Garon, and Thierry Langer.
\newblock Applications of the pharmacophore concept in natural product inspired
  drug design.
\newblock {\em Molecular Informatics}, 39(11):2000059, 2020.

\bibitem{Gilson:Given:Bush:McCammon:97}
M.~K. Gilson, J.~A. Given, B.~L. Bush, and J.~A. McCammon.
\newblock The statistical-thermodynamic basis for computation of binding
  affinities: A critical review.
\newblock {\em Biophys. J.}, 72:1047--1069, 1997.

\bibitem{Gallicchio2011adv}
Emilio Gallicchio and Ronald~M Levy.
\newblock Recent theoretical and computational advances for modeling
  protein-ligand binding affinities.
\newblock {\em Adv. Prot. Chem. Struct. Biol.}, 85:27--80, 2011.

\bibitem{cournia2020rigorous}
Zoe Cournia, Bryce~K Allen, Thijs Beuming, David~A Pearlman, Brian~K Radak, and
  Woody Sherman.
\newblock Rigorous free energy simulations in virtual screening.
\newblock {\em Journal of Chemical Information and Modeling}, 2020.

\bibitem{Gallicchio2021binding}
Emilio Gallicchio.
\newblock Free energy-based computational methods for the study of
  protein-peptide binding equilibria.
\newblock In Thomas Simonson, editor, {\em Computational Peptide Science:
  Methods and Protocols}, Methods in Molecular Biology. Springer Nature, 2021.

\bibitem{Gallicchio2010}
Emilio Gallicchio, Mauro Lapelosa, and Ronald~M. Levy.
\newblock Binding energy distribution analysis method ({BEDAM}) for estimation
  of protein-ligand binding affinities.
\newblock {\em J. Chem. Theory Comput.}, 6:2961--2977, 2010.

\bibitem{Tan2012}
Zhiqiang Tan, Emilio Gallicchio, Mauro Lapelosa, and Ronald~M. Levy.
\newblock Theory of binless multi-state free energy estimation with
  applications to protein-ligand binding.
\newblock {\em J. Chem. Phys.}, 136:144102, 2012.

\bibitem{Boresch:Karplus:2003}
S~Boresch, F~Tettinger, M~Leitgeb, and M~Karplus.
\newblock Absolute binding free energies: A quantitative approach for their
  calculation.
\newblock {\em J. Phys. Chem. B}, {107}:{9535--9551}, {2003}.

\bibitem{pronk2013gromacs}
Sander Pronk, Szil{\'a}rd P{\'a}ll, Roland Schulz, Per Larsson, P{\"a}r
  Bjelkmar, Rossen Apostolov, Michael~R Shirts, Jeremy~C Smith, Peter~M Kasson,
  David van~der Spoel, Berk Hess, and Erik Lindahl.
\newblock Gromacs 4.5: a high-throughput and highly parallel open source
  molecular simulation toolkit.
\newblock {\em Bioinformatics}, 29:845--854, 2013.

\bibitem{khuttan2021single}
S~Khuttan, Solmaz Azimi, Joe~Z Wu, and E~Gallicchio.
\newblock Alchemical transformations for concerted hydration free energy
  estimation with explicit solvation.
\newblock {\em J. Chem. Phys}, 154:054103, 2021.

\bibitem{eastman2017openmm}
Peter Eastman, Jason Swails, John~D Chodera, Robert~T McGibbon, Yutong Zhao,
  Kyle~A Beauchamp, Lee-Ping Wang, Andrew~C Simmonett, Matthew~P Harrigan,
  Chaya~D Stern, et~al.
\newblock Openmm 7: Rapid development of high performance algorithms for
  molecular dynamics.
\newblock {\em PLoS Comp. Bio.}, 13(7):e1005659, 2017.

\bibitem{gallicchio2015asynchronous}
Emilio Gallicchio, Junchao Xia, William~F Flynn, Baofeng Zhang, Sade
  Samlalsingh, Ahmet Mentes, and Ronald~M Levy.
\newblock Asynchronous replica exchange software for grid and heterogeneous
  computing.
\newblock {\em Computer Physics Communications}, 196:236--246, 2015.

\bibitem{rizzi2018overview}
Andrea Rizzi, Steven Murkli, John~N McNeill, Wei Yao, Matthew Sullivan,
  Michael~K Gilson, Michael~W Chiu, Lyle Isaacs, Bruce~C Gibb, David~L Mobley,
  et~al.
\newblock Overview of the sampl6 host--guest binding affinity prediction
  challenge.
\newblock {\em J. Comp.-Aid. Mol. Des.}, 32(10):937--963, 2018.

\bibitem{shi2021amoeba}
Yuanjun Shi, Marie~L Laury, Zhi Wang, and Jay~W Ponder.
\newblock Amoeba binding free energies for the sampl7 trimertrip host--guest
  challenge.
\newblock {\em J. Comp.-Aid. Mol. Des.}, 35(1):79--93, 2021.

\bibitem{Azimi2021RBFE}
Solmaz Azimi, Sheenam Khuttan, Joe~Z. Wu, Rajat Pal, and Emilio Gallicchio.
\newblock Relative binding free energy calculations for ligands with diverse
  scaffolds with the alchemical transfer method.
\newblock {\em ArXiv Preprint}, XXX:XXX--XXX, 2021.

\end{thebibliography}

\end{document}